\documentclass{article}
\usepackage{spconf,amsmath,graphicx,xcolor}
\usepackage{multirow}
\usepackage{arydshln}

\usepackage{soul}

\title{Improving Noise Robustness for Spoken Content Retrieval using semi-supervised ASR and N-best transcripts for BERT-based \\ ranking models}

\name{Yasufumi Moriya, Gareth. J. F. Jones}
\address{ADAPT Centre, School of Computing, Dublin City University, Dublin 9, Ireland}
\copyrightnotice{978-1-6654-7189-3/22/\$31.00~\copyright2023 European Union}

\begin{document}
%
\maketitle
\begin{abstract}
BERT-based re-ranking and dense retrieval (DR) systems have been shown to improve search effectiveness for spoken content retrieval (SCR). However, both methods can still show a reduction in effectiveness when using ASR transcripts in comparison to accurate manual transcripts. 
We find that a known-item search task on the How2 dataset of spoken instruction videos shows a reduction in mean reciprocal rank (MRR) scores of 10-14\%.  As a potential method to reduce this disparity, we investigate the use of semi-supervised ASR transcripts and N-best ASR transcripts to mitigate ASR errors for spoken search using BERT-based ranking. Semi-supervised ASR transcripts brought 2-5.5\% MRR improvements over standard ASR transcripts and our N-best early fusion methods for BERT DR systems improved MRR by 3-4\%. Combining semi-supervised transcripts with N-best early fusion for BERT DR reduced the MRR gap in search effectiveness between manual and ASR transcripts by more than 50\% from 14.32\% to 6.58\%.


\end{abstract}
\section{Introduction}
\label{sec:intro}

The growing archives of spoken multimedia content, such as those arising in video streaming services and podcasts, are increasing demands for effective spoken content retrieval (SCR) systems. Unlike textual documents, spoken documents are not directly searchable based on their content. For this reason, spoken multimedia content is often searched only based on user-created summaries and metadata. The effectiveness of this approach relies on availability of meaningful summaries and metadata for the documents. Further, since many spoken content items are long, SCR systems should ideally provide users with more sophisticated, finer granularity search functionalities based on the content itself. 
Such functionalities can only be provided if transcripts of the spoken contents of the documents are available.

The high cost of creating manual transcripts and the large amounts of content mean that transcripts are generally made
using automatic speech recognition (ASR). However, transcription errors from ASR can impact on search effectiveness. While information retrieval (IR) has been 
shown to generally be robust to moderate
word error rates (WERs), Larson and Jones report that WERs greater than about 30\% can significantly impact on search reliability \cite{larson2012}. The accuracy of ASR systems continues to improve, with very low WERs 
reported for some tasks \cite{wang2020transformer,hsu2021hubert}. However, WERs can still though be high when there is domain mismatch between the training of the ASR system and the data to be transcribed \cite{moriya2021augmenting}, 
speech is informal or 
acoustic noise is present in audio data \cite{wang2020transformer}. In many settings, there is no control of domains or audio quality of the data, e.g., user-generated data of video streaming services or Podcasts, unlike audio corpora used in laboratory settings. Thus, even a state-of-the-art ASR system can still suffer high transcription WERs. This motivates us to explore 
development of methods to improve
noise robustness of SCR systems.

Transformer-based IR models have recently emerged as a highly effective alternative to traditional models such as BM25 \cite{robertson2009}. These new models can largely be classified into two types: re-ranking and dense retrieval (DR). The BERT-based re-ranking model, namely MonoBERT \cite{nogueira2019passage} directly takes a query-document pair to compute its relevance score. The drawback of MonoBERT and its successors is that transformer-based inference is very slow and hardware requirements can make them impractical for use in practice \cite{xiong2021approximate}. 
In DR, the transformer-based model is used to encode documents into dense document vectors separately from queries. At search time, only the query needs to be encoded into vector representations. Then, any vector-based 
similarity measure, such as nearest neighbour (NN), can be used to compute query-document relevance scores. This approach is much faster to run than re-ranking \cite{xiong2021approximate}. 

The contributions of this paper are as follows. (i) We investigate BERT-based re-ranking and DR for a known-item search task of spoken instruction videos. We observe that, despite the use of BERT-based retrieval systems, the gap in search effectiveness between manual and ASR transcripts was around 10-14\%. (ii) We augment an ASR system using semi-supervised training for domain adaptation and examine the impact of improved ASR transcripts on the BERT-based ranking systems. The use of semi-supervised ASR transcripts brought 2-5.5\% improvement of the MRR score. (iii) We propose the use of N-best ASR transcripts for BERT DR systems and observe that our proposed early fusion approach brought a gain in the MRR score by around 3-4\%. Combining these two methods gives an average improvement in MRR of 6.5\%.

The reminder of the paper is organised as follows. The next section gives 
a brief review of BERT-based IR and SCR, followed by details of the models used in this investigation. We then propose our early and late fusion extensions using ASR N-best transcripts for a BERT DR system. Experimental results and analysis of the BERT-based ranking models are then presented, followed by conclusions and future work.

\section{Relevance to Prior Work}
\label{sec:relevance}

Mainstream studies of spoken document retrieval (SDR) began with 
the Text REtrieval Conference (TREC) \cite{garofolo2000trec}. Studies have continued with exploration of SCR in activities such as those at 
MediaEval 
\cite{eskevich2014search} and the Podcast retrieval track at TREC 2020 and 2021 \cite{jones2020trec,jones2021trec}. Another line of recent studies on SCR
is 
described in \cite{lin2019enhanced,jiang2020spoken} 
, which 
investigate the use of BERT re-ranking on retrieval of 
Mandarin news stories. 

A number of retrieval models have been introduced over the years, among the most consistently effective is the BM25 probabilistic model \cite{robertson2009}, which has been applied successfully in many SCR tasks including \cite{eskevich2014search,jones2020trec}. Recent years have seen the introduction of neural IR models, including \cite{dai2018convolutional}. However, more recently a new class of retrieval models using the pre-trained transformer architecture (BERT) \cite{devlin-etal-2019-bert} has 
been introduced \cite{lin2020pretrained}. These models have been shown to be better than the classic BM25 model \cite{robertson2009} and non-BERT neural models, such as Conv-KNRM \cite{dai2018convolutional}. 
These BERT-based search models can be classified into two categories: BERT-based re-ranking \cite{nogueira2019passage} and dense retrieval (DR) \cite{xiong2021approximate,karpukhin2020dense}. While BERT-based re-ranking takes as input a query-document pair and directly produces a relevance score, DR encodes documents and queries into dense vectors independently and applies 
approximate nearest neighbour (ANN) search 
to 
these
vectors.

N-best ASR hypotheses have been used as input to
the BERT model for Spoken Language Understanding as a form of lattices \cite{Huang2019adapting}, confusion networks \cite{liu2020jointly} and plain texts \cite{ganesan2021nbest}. Ganesan et al. conclude in \cite{ganesan2021nbest} 
that text format input was more suitable for BERT models than alternative formats since BERT models are pre-trained on plain texts. N-best hypotheses have also been explored for ASR error correction \cite{zhu2021improving}. 

Our paper differs from 
existing work on BERT-based ranking systems for SCR tasks as follows. (i) We experiment with both BERT re-ranking and BERT DR systems on manual transcripts and ASR transcripts. The TREC Podcasts Tracks \cite{jones2020trec,jones2021trec} have seen the use of BERT re-ranking and DR models. However, manual transcripts were not available for system evaluation. 
BERT re-ranking was investigated in \cite{lin2019enhanced,jiang2020spoken}, while the BERT DR model was not applied to their spoken broadcast retrieval. (ii) We employ semi-supervised ASR transcripts to 
the SCR setting, 
where a domain-adapted ASR system is not available. Existing literature \cite{larson2012,garofolo2000trec} suggests 
a near-linear relationship between transcription WERs and search scores. Our investigation into the use of semi-supervised ASR transcripts demonstrates that BERT re-ranking and DR models benefit from improved transcripts using a semi-supervised ASR system. (iii) We propose the use of ASR N-best transcripts for BERT DR systems. To the best of our knowledge, BERT ranking models using ASR N-best for SCR has 
not been investigated previously. We observe that both early and late fusion approaches for BERT DR mitigate errors in ASR transcripts for the SCR task.

\section{BERT-based Retrieval Models}
\label{background}

In this section we provide details of BERT-based re-ranking and DR. We first present the models, and then describe how they are trained.

\subsection{BERT for retrieval systems}
\label{bert_retrieval}

\begin{figure}[t]
  \centering
  \includegraphics[width=\linewidth]{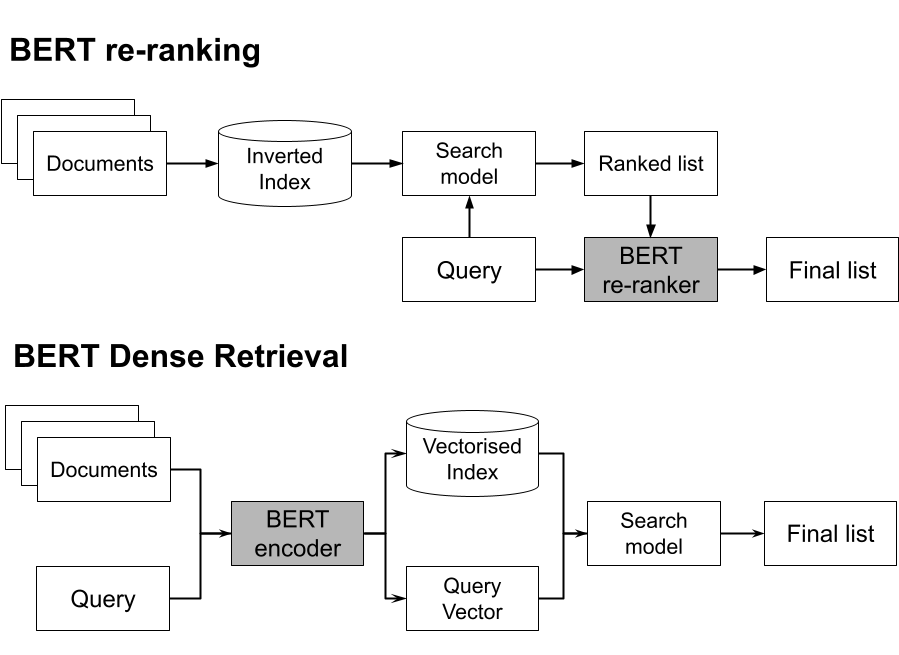}
  \caption{Schematic diagram of a BERT-based re-ranking system and a BERT-based dense retrieval (DR) system.}
  \label{fig01:bert_retrieval}
\end{figure}

Figure~\ref{fig01:bert_retrieval} shows the structure of a BERT re-ranking system and a BERT DR system. The BERT re-ranker takes a ranked document list from an initial search model and re-ranks it using the BERT model, while the DR system uses BERT to transform documents and a query into dense vectors and another search model to obtain the final ranked list of documents.

\vspace{0.5ex}
\noindent
{\em BERT Re-ranker \/} 
The BERT model re-ranks a ranked document list retrieved from a document archive using a lightweight search model such as BM25. The first BERT model used for re-ranking was MonoBERT \cite{nogueira2019passage}. Suppose a query consists of $N$ tokens and a document $M$ tokens, the input to the BERT model can be expressed as $[CLS] q_1, ..., q_N [SEP]$ $d_1, ..., d_M [SEP]$, where $[CLS]$ and $[SEP]$ are special tokens of BERT. A relevance score of the query-document pair is computed by feeding an output vector corresponding to the $[CLS]$ token to a fully-connected layer. 

\vspace{0.5ex}
\noindent
{\em BERT Dense Retrieval \/} 
A BERT model is used to encode a query and documents into vector representations. While two BERT models were used by Karpukhin et al. \cite{karpukhin2020dense} to encode a query and documents independently, Xiong et al. \cite{xiong2021approximate} applied a single BERT model to encode a query and documents (Dual encoder). In this work, we used the dual encoder framework for our
DR system. A representation of a query and a document is a vector corresponding to the $[CLS]$ token preceding query tokens or document tokens. In contrast
to the re-ranker BERT output which is a relevance score, a dense vector is a representation of a document or a query in a fixed dimensional space, and the final relevance score is computed by taking a similarity score of the query and document vectors.

A similarity score can be computed using a dot product or cosine similarity. This enables an efficient retrieval algorithm, such as approximate nearest neighbour, to be used to create a final list of documents according to query-document similarity scores \cite{johnson2019billion}. In this work, we use a dot product for 
similarity scoring following \cite{xiong2021approximate}. 

\vspace{0.5ex}
\noindent
{\em BERT re-ranking vs BERT dense retrieval \/} 
BERT re-ranking has generally been found to produce better retrieval metrics than the BERT DR \cite{lin2020pretrained}. However, Xiong et al. \cite{xiong2021approximate} demonstrated that using it
can be 100 times slower than DR at search time. Due to its inference speed, BERT re-ranking is often applied to only the $k$ top documents returned by an
initial search, while BERT DR needs only to encode a query at search time and can be applied to the whole document collection.

\subsection{BERT training for retrieval}

\vspace{0.5ex}
\noindent
{\em BERT Re-ranking \/} 
The BERT re-ranking model is trained using binary cross-entropy loss \cite{nogueira2019passage}. Specifically, given a query and a relevant document pair, the BERT model is trained to produce an output of 1, while it
is trained to produce 0 for a query and a non-relevant document. This can be expressed as
\begin{equation}
    L = -\log(f(q_i, d_i^+; \theta)) - \sum_{j=1}^{J} \log(1 - f(q_i, d_{ij}^-; \theta))
\end{equation}
where $f(q_i, d_i^+; \theta)$ is the score from the BERT model given the $i$th query-document pair and $f(q_i, d^-_{ij}; \theta)$ the score from the BERT model given the $j$th negative query-document pair corresponding to the $i$th train example.

\vspace{0.5ex}
\noindent
{\em BERT Dense Retrieval \/} 
The BERT DR model is trained using the negative log likelihood \cite{xiong2021approximate,karpukhin2020dense}. Specifically, the model is trained to maximise a similarity score of a pair of a query and a relevant document and to minimise a similarity score of a pair of query and a non-relevant document. Formally, this is:
\begin{align} 
    s(q, d) &= sim(g(q; \theta), g(d; \theta))\\
    L &= -\log \frac{\exp(s(q_i, d_i^+))}{\exp(s(q_i, d_i^+)) + \sum^J_{j=1} \exp(s(q_i, d_{ij}^-))}
\end{align}
where $g(q; \theta)$ is a function to obtain a query dense vector corresponding to the BERT output vector of the [CLS] token and $g(d; \theta)$ its document counterpart. $sim()$ is a function to compute a similarity score of the two vectors. Similar to the cross entropy loss used for the BERT re-ranker, this loss is obtained for the $i$th query-document pair and corresponding $j$ negative query-document pairs.

To obtain negative examples for each true query-document pair, the BM25 model is often used to decide the initial negative examples. Xiong et al. \cite{xiong2021approximate} proposed an improved algorithm to select negative examples named Approximate nearest neighbour Negative Contrastive Learning (ANCE). This algorithm begins with BERT training on negative examples chosen from the output of BM25 document ranking, and updates negative examples using the BERT model being trained at every specified checkpoint. Xiong et al. \cite{xiong2021approximate} provides theoretical grounds for
ANCE and empirical results demonstrating that ANCE training produces better search results and faster model convergence. While their original paper proposes asynchronous updates of negative examples where inference for ANCE and model training run in parallel, we stopped model training at the end of every training epoch and ran ANCE updates to resume model training on updated negative examples.

\section{N-best Fusion for BERT Dense Retrieval}
\label{nbest}

As outlined 
in Section~\ref{sec:relevance}, using ASR N-best transcripts has been effective in SLU and ASR error correction \cite{ganesan2021nbest,zhu2021improving}. In this paper, we propose two N-best fusion approaches for BERT DR. The first approach fuses N-best transcripts before feeding text to BERT (early fusion). The second approach creates N indexes for N-best transcripts and combines scores from each of the indexes to compute the final score for each document (late fusion). 

\subsection{N-best early fusion}

The input sequence length of the BERT model is often limited to 512 tokens with tokens exceeding this maximum length being truncated \cite{devlin-etal-2019-bert}. For the task of SLU, user utterances are generally very short and N-best transcripts can be concatenated with the [SEP] token \cite{ganesan2021nbest}. However, in case of SCR, document lengths are generally much longer and N-best document representations cannot be simply concatenated with the [SEP] token. This issue can be resolved by the alignment approach introduced in \cite{zhu2021improving}, and taking the average of $N$ word embeddings for $i$th token to compress $N$ documents into a single document representation. 

\begin{figure}[t]
  \centering
  \includegraphics[width=\linewidth]{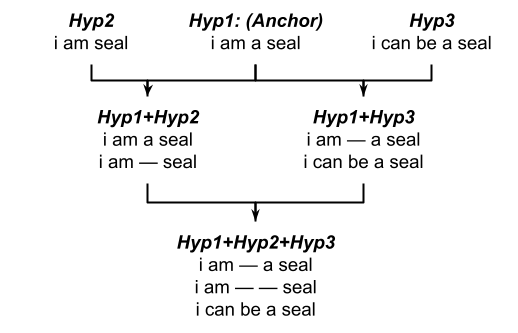}
  \setlength{\belowcaptionskip}{-15pt}
  \vspace{-7mm}
  \caption{Example of N-best alignment to create BERT input text.}
  \vspace{-5mm}
  \label{fig02:bert_early}
\end{figure}

Figure~\ref{fig02:bert_early} shows an example of aligning 3-best transcripts. The 1-best transcript is used as an anchor transcript and aligned with the other two transcripts. The aligned transcripts are further combined to match the length across N-best transcripts. As described in 
\cite{zhu2021improving}, ``---'' is used as a special token to alleviate a length mismatch between transcripts. This token is added to one of the word embedding vectors. The input of the BERT model is often sub-word tokens \cite{devlin-etal-2019-bert}. It is important to run alignment of transcripts \emph{after} tokenisation is applied to texts. If this order is reversed, aligned transcripts can create a length mismatch after tokenisation. Once the N-best transcripts have been aligned, the tokens are transformed into word embeddings with 
the $i$th position of the aligned tokens 
combined by taking the average. For example, in Figure~\ref{fig02:bert_early}, the first position of the three transcripts are
all ``i''. 
This results in the averaged embeddings of ``i''. This could be extended by concatenating embeddings and feeding concatenated embeddings to a linear layer, as 
in \cite{zhu2021improving}, but the concatenated vector size can be large and we did not apply this method in this paper.

\vspace{-3mm}

\subsection{N-best late fusion}

Our N-best late fusion approach to BERT DR involved creation of an $N$ index for each of the N-best transcripts. 
The BERT model encodes a query into a vector and search is carried out for all $N$ indexes. The final result is calculated by summing the scores for each document from each index, and ranking in descending order of 
summed scores. While a
drawback of this approach is creating $N$ indexes, ANN search over $N$ indexes could 
be run in parallel using $N$ CPUs after computing a query vector to maintain operational speed.

\section{Experimental Investigation}
\label{experiments}

In this section we present our experimental investigation. We 
first 
describe our experimental setup. This includes details of the data used for our experiments, training and hyper-parameter details of BERT re-ranking and DR models, and creation of standard ASR transcripts and semi-supervised ASR transcripts. We then
present experimental results of a known-item search task using a corpus of spoken instruction videos. The results show search scores of applying BERT ranking systems to different types of transcripts, inference time of BERT re-ranking and DR systems, and the use of an N-best extension for BERT DR systems.

\subsection{Experimental Setup}

{\em Test Collection and Evaluation} 
Our experimental investigation was carried out using the How2 dataset \cite{sanabria18how2}. How2 consists of 19,770 videos with manual transcripts and metadata including video titles. For our experiments we used How2 video titles as search queries to retrieve the associated video to form a known-item search task where each query has one relevant target document. We randomly selected 500 
of the 19,770 video titles as evaluation queries. Experiments used the standard known-item search of mean reciprocal rank (MRR), defined as follows:
\begin{equation}
\label{02:mrr}
    MRR = \frac{1}{N}\sum^N_{i=1}\frac{1}{rank_i}
\end{equation}
where $N$ is the number of user queries and $rank_i$ is the rank of the relevant document in the retrieval list for the $i$th query. In the How2 corpus, only around 2.5\% of the collection used exceeds the length limit of BERT. Thus, our results are unlikely 
to be affected by the BERT length limit. Examination of any 
impact of this 
issue 
and strategies to address it 
could form 
future work.

\vspace{0.5ex}

\noindent {\em BERT Retrieval Model Data Setup\/} For the re-ranking BERT model, an additional 500 validation queries were randomly selected. For each validation query, one positive example and one negative example from the top of the BM25 ranked document list were chosen (if the top document is the positive example, the second ranked document was chosen) to compute validation loss using the cross entropy. For the BERT DR model, 500 validation queries were randomly chosen, and for each query, 1 relevant document and 99 irrelevant documents from the BM25 model were ranked by the BERT encoder to compute the mean reciprocal rank of the relevant document. 

\vspace{0.5ex}

\noindent {\em ASR Transcripts\/} To obtain ASR transcripts of How2 audio data, a
Kaldi ``chain'' system was developed using the LibriSpeech corpus \cite{Panayotov2015librispeech,Hadian2018endtoend}. This is a traditional ASR system integrating an acoustic model with a hidden Markov model. This
system, referred to as ``standard ASR'', trained on LibriSpeech created a domain mismatch scenario to recognise How2 audio and produced an overall WER of 31.1\%. We used the ``standard ASR'' system to generate decode lattices and 1-best transcripts of the How2 train data. Combined with LibriSpeech data, the decode lattices were used to train a domain-adapted semi-supervised acoustic model, while 1-best transcripts were used for training of a semi-supervised N-gram for decoding and a re-scoring recurrent neural network language model \cite{moriya2021augmenting,Manohar2018,xu2018}. This system, referred to as ``semi-supervised ASR'' 
produced an overall WER of 23.95\%. Further, we generated lattice oracle transcripts using the ``standard ASR'' system (i.e., the most correct hypotheses present in lattices with respect to manual transcripts), which produced an overall WER of 11.18\%. For the N-best experiments, the ``standard ASR'' and ``semi-supervised ASR'' systems were used to generate \{2, 5, 10, 20\}-best transcripts to develop the N-best version of the BERT DR system. 

\vspace{0.5ex}

\noindent {\em Retrieval Models\/} The BM25 model used was the Python version of Lucene with $k1$ 1.2 and $b$ 0.75\footnote{https://lucene.apache.org/}. The re-rank system took as input the top 1,000 documents returned for each query.
The pre-trained BERT model used was ``bert-base-uncased'' and its corresponding tokeniser from HuggingFace \cite{wolf2020transformers}. For the DR system, after the whole document collection had been encoded by the BERT model, Faiss software \cite{johnson2019billion} was used to run similarity search over a query vector and an index of document vectors. 

The hyper-parameter settings for BERT model training were as follows. The re-ranker BERT was trained with 20 epochs and used the model checkpoint which produced the best validation score. Model training was generally converged by epoch 15. The best initial learning rate was empirically selected, and 
was $3 \times 10^{-7}$. The DR BERT was converged by 10 epochs as reported by \cite{xiong2021approximate}. The best initial learning rate found was $1 \times 10^{-5}$. Both re-ranker BERT and DR BERT used 20 
negative examples.
Our BERT models were trained using Quadro RTX 6000. The BERT models could 
potentially be improved by using more negative examples, but the number of negative examples 20 was its memory limit. Increasing the negative examples using a memory queue could form future work.
For training of the re-ranker and DR BERT, the AdamW optimizer was used with a
weight decay of $5 \times 10^{-5}$ to avoid overfitting.

\subsection{MRR scores using lattice oracle, manual, standard and semi-supervised ASR transcripts}
\label{subsec:results_transcripts}

Table~\ref{table_transcripts_how2} presents MRR scores for
BM25, BERT re-ranking and BERT DR systems using manual (man), lattice oracle (oracle), semi-supervised ASR (semi ASR) and standard ASR (std ASR) transcripts. We use the paired 2-tailed t-test to compare distributions of MRR scores from BM25 and other BERT-based models.

\vspace{0.2ex}
\noindent {\em Comparison of BERT ranking systems\/}
The largest MRR gain was obtained using the BERT DR system trained with ANCE. This
improved the MRR scores by around 26\% over the BM25 model on manual transcripts, and by
23\% on standard ASR transcripts, indicating that transcription errors affected the score. The BERT re-ranking system improved the MRR scores by around 16-18\% over the BM25 model for all transcript types.

\vspace{0.5ex}
\noindent {\em Comparison of searching over different transcripts\/} Despite the 
improvement in MRR obtained by BERT re-ranking and DR systems, the gap remains 
between manual and ASR transcript scores.
Searching over standard ASR transcripts resulted in around 11\% worse MRR scores than manual transcripts with BM25 and BERT re-ranking. BERT DR systems were more affected by transcription errors showing around 15\% worse MRR scores when using standard ASR transcripts. Although semi-supervised transcripts had a small impact on the improvement of MRR score by the BM25 model, the BERT re-ranking system improved the MRR score by 2\% and the BERT DR systems improved the score by 5.5\% without ANCE training and by around 4\% with ANCE training. The oracle experiments, however, show that the gap in the MRR scores between manual transcripts and ASR transcripts could potentially be further reduced. In comparison to searching over manual transcripts, all of the models including BM25 produced only 3-4\% worse MRR scores when searching over lattice oracle transcripts.

\begin{table}[t]
\centering
\caption{MRR results of using different transcripts for How2 known-item search evaluation. All of BERT results are statistically significant over BM25 with $p < 0.05$ denoted by **.}
\begin{tabular}{ccccc}
\hline
\multirow{3}{*}{model} & \multicolumn{4}{c}{MRR} \\ 
  & \multicolumn{4}{c}{eval data}\\\cline{2-5}
  & man & oracle & semi ASR & std ASR \\ \hline
BM25   &  40.27 & 36.02 & 30.85 & 29.37 \\\hline
rerank & 57.32** & 52.46** & 48.69** & 46.37** \\
DR & 50.61** & 45.96** & 40.38** & 34.81** \\
DR-ANCE &\textbf{66.73}** & \textbf{63.88}** & \textbf{56.22}** & \textbf{52.41}** \\\hline
\end{tabular}
\label{table_transcripts_how2}
\vspace{-5mm}
\end{table}

\vspace{0.5ex}
\noindent {\em Comparison of search run time\/}
Similar to \cite{xiong2021approximate}, we compare the inference time of 100 documents per query. Search operations of BM25 were measured using a CPU, while inference of BERT re-ranking was performed using a GPU. 
BERT DR search run time 
is the 
total 
for encoding the query using a GPU and ANN operations using a CPU. For fair comparison, all of the search runs were performed on the same computation node of our high performance computing system. We observe that average run time of BM25 was 40 msecs,
BERT re-ranking 1.84 secs,
and BERT DR 18 msecs,
respectively. This result agrees with \cite{xiong2021approximate}, where
BERT re-ranking was 
roughly 100x slower than BERT DR at search time and demonstrates a practical advantage of the BERT DR system.

\begin{table*}[t]
\centering
\caption{MRR results of BERT-DR models with and without ANCE training using the N-best extension over standard ASR and semi-supervised ASR transcripts. \% change shows relative improvement/regression of MRR over corresponding 1-best models. Asterisks * and ** denote statistical significance over the baseline 1-best system with $p < 0.05$ and $p < 0.01$, respectively.}
\begin{tabular}{ccccccc}
\hline

\multirow{3}{*}{model}  &  \multicolumn{5}{c}{MRR} & \multirow{3}{*}{\% change}\\
 &  \multicolumn{5}{c}{N-size} &  \\\cline{2-6}
 & 1 & 2 & 5 & 10 & 20 \\\hline
\textit{standard ASR} & & & & \\ \cdashline{2-7}
DR-early &\multirow{2}{*}{34.81}&  \textbf{35.42} & 32.94 & 33.71 & 34.24 & +1.75 (2-best)\\
DR-late & & 35.49 & 36.94** & 37.55** & \textbf{38.20}** & +9.74 (20-best) \\ \cdashline{2-7}
DR-ANCE-early & \multirow{2}{*}{52.41}& 51.77 & 52.14& 52.75 & \textbf{54.45} &+3.49 (20-best) \\
DR-ANCE-late & & 53.29 & 54.88** & 54.86** & \textbf{55.40}** & +5.71(20-best)\\\hline
\textit{semi-sup ASR} & & & & \\ \cdashline{2-7}
DR-early & \multirow{2}{*}{40.38} & 39.40 & 38.21 & 37.91 & 38.72 & -2.43 (2-best) \\
DR-late & & 40.05 & 40.21 & \textbf{41.20}  & 40.93 & +2.03 (10-best) \\ \cdashline{2-7}
DR-ANCE-early & \multirow{2}{*}{56.22} & 57.26 & 57.27 & 57.92 & \textbf{60.15}** & +6.99 (20-best) \\
DR-ANCE-late & & 57.79* & 58.19** & 58.79** & \textbf{59.20}** & +5.30 (20-best)\\\hline
\end{tabular}
\label{table_nbest}
\end{table*}

\subsection{MRR scores using ASR N-best for BERT DR}
\label{subsec:results_nbest}

We next present the MRR scores of our proposed ASR N-best extension for BERT DR. Table~\ref{table_nbest} summarises the MRR scores of BERT DR using \{2,5,10,20\}-best transcripts either from the standard ASR system or from the semi-supervised ASR system. MRR scores of N-size 1 correspond to the MRR scores from
row 3 and row 4 and semi ASR and std ASR columns in Table~\ref{table_transcripts_how2}. Bold face indicates 
MRR scores of each BERT DR system which are higher than the baseline 1-best result and highest among other N sizes. A paired 2-tailed t-test was again carried out to compare MRR distributions from the 1-best BERT model and other N-best models.

The 20-best early and late fusion over semi-supervised transcripts using ANCE BERT produced MRR of 60.15\% and 59.20\%, respectively (row 3 and 4 of \textit{semi-sup ASR} in Table~\ref{table_nbest}). Considering that the 1-best BERT DR over standard ASR transcripts produced MRR of 52.41\% and the MRR gap between standard ASR and manual transcripts was 14.32\% ($66.73 - 52.41$), the MRR score gap was reduced to 6.58\% ($66.73-60.15$) using
the 20-best early fusion approach. This reduction 
represents a more than 50\% relative improvement.

\vspace{0.5ex}
\noindent {\em Early vs late fusion\/} While the early fusion approach produced inconsistent improvement of MRR with different N-sizes, the late fusion approach consistently benefited from increase in N sizes. Despite the highest MRR score of 60.15\% obtained from early fusion of 20-best transcripts for BERT DR with ANCE training, when the early fusion was applied to BERT DR without ANCE training, the MRR scores did not improve with
increase in the value of N.
While statistical significance over the 1-best system was observed over semi-supervised ASR transcripts (row 3 \textit{semi-sup ASR}), the MRR score of 20-best over the standard ASR transcripts did not show statistical significance (row 3 \textit{standard ASR}). In contrast, the late fusion approach brought consistent improvement over 1-best systems. The highest MRR score observed was 59.20 using 20-best semi-supervised ASR transcripts for BERT DR with ANCE training (row 4 \textit{semi-sup ASR}). Generally, 20-best transcripts benefited the late fusion approach the most, except when 
BERT DR 
without ANCE training was applied to semi-supervised ASR transcripts (row 2 \textit{semi-sup ASR}). 

In further analysis we examined the average MRR improvement and regression per query for 
the 20-best extension to ANCE BERT DR systems. The early fusion approach brought an
average 
improvement in MRR of 0.32 for
147 out of 500 evaluation queries over the 1-best system, while 123 out of these 500 evaluation queries had a lower MRR score with average 0.22. MRR scores of the other 240 queries were unchanged. The late fusion approach with
20-best ANCE BERT improved the MRR scores of 135 queries with by an 
average 0.17, while reducing
scores of 86 queries by 
an average 
of 
0.10. MRR scores of the other 279 queries were unchanged by the late fusion approach. These results demonstrate that the early fusion approach introduces more radical changes than late fusion, in computation of document representations over 1-best systems. 

\section{Conclusions and Future Work}

We explored BERT re-ranking and DR systems for a known-item SCR task. 
Despite improved search effectiveness using the BERT ranking model, MRR scores from BERT re-ranking and DR systems were 10-14\% lower when 
using ASR 
than manual transcripts. We investigated the use of semi-supervised ASR transcripts and an N-best extension for BERT DR. 
Use of semi-supervised transcripts improved MRR by 2-5.5\%, while 
use of 20-best transcripts brought a gain in 3-4\%. Combining them 
gave an average improvement 
of 6.5\%.

Future work includes the use of decode lattices for BERT retrieval. As seen in Table~\ref{table_transcripts_how2}, applying BERT ranking models to lattice oracle transcripts further closed the MRR gap in 3-5\% MRR to manual transcripts. This motivates development of BERT ranking which can effectively take as input decode lattices to overcome ASR transcription errors.

%

\noindent {\em Acknowledgement\/}: This work was supported by Science Foundation Ireland as part of the ADAPT Centre (Grant 13\//RC\//2106) at Dublin City University.

\bibliographystyle{IEEEbib}
\bibliography{strings,refs}

\end{document}